\documentclass{article}
\usepackage{spconf,amsmath,graphicx}
\usepackage{MnSymbol}
\usepackage{amsthm}
\usepackage{amsfonts}
\usepackage[font=small,skip=.5pt]{caption}
\usepackage{algorithm}
\usepackage{algorithmicx,algpseudocode}
\usepackage{xspace}

\usepackage[inline]{enumitem}
\newlist{inlinelist}{enumerate*}{1}
\setlist*[inlinelist,1]{label=\roman*),itemjoin={{, }},itemjoin*={{, and }}}

\usepackage{listings}
\usepackage[labelformat=simple]{subcaption}
\DeclareCaptionLabelSeparator{periodspace}{.\quad}
\usepackage{multirow}
\usepackage{pbox}
\usepackage{url}
\usepackage{comment}
\usepackage{hyperref}
\usepackage{color}

\usepackage{bm}

\newcommand{\fv}{{\bf f}}
\newcommand{\gv}{{\bf g}}

\newcommand{\uv}{{\bf u}}

\newcommand{\xv}{{\bf x}}

\newcommand{\Dm}{{\bf D}}

\newcommand{\Gm}{{\bf G}}
\newcommand{\Hm}{{\bf H}}
\newcommand{\Id}{{\bf I}}

\newcommand{\Lm}{{\bf L}}

\newcommand{\Qm}{{\bf Q}}

\newcommand{\Sm}{{\bf S}}

\newcommand{\Um}{{\bf U}}
\newcommand{\Wm}{{\bf W}}

\newcommand{\Ec}{{\cal E}}

\newcommand{\Gc}{{\cal G}}

\newcommand{\Vc}{{\cal V}}

\newcommand{\deltav}{\hbox{\boldmath$\delta$}}

\newcommand{\Lambdam}{\hbox{\boldmath$\Lambda$}}

\DeclareMathOperator{\dist}{d}

\title{What's in a frequency: new tools for graph Fourier Transform visualization}

\name{Benjamin Girault and Antonio Ortega
\address{Signal and Image Processing Institute\\Department of Electrical and Computer Engineering, \\
University of Southern California, Los Angeles, USA}
}

\def\matlab{\texttt{Matlab}\xspace}
\def\grasp{\texttt{GraSP}\xspace}

\begin{document}
\ninept
\maketitle
\begin{abstract}
Recent progress in graph signal processing (GSP) has addressed a number of problems, including sampling and filtering. Proposed methods have focused on generic graphs and defined signals with certain characteristics, \textit{e.g.}, bandlimited signals, based on the graph Fourier transform (GFT). However, the effect of GFT properties (\textit{e.g.}, vertex localization) on the behavior of such methods is not as well understood. In this paper, we propose novel GFT visualization tools and provide some examples to illustrate certain GFT properties and their impact on sampling or wavelet transforms.
\end{abstract}
\begin{keywords}
Graph signal processing, graph Fourier transform, visualization, filtering, sampling.
\end{keywords}
\section{Introduction}
\label{sec:intro}

Significant recent progress has been made in developing graph signal processing (GSP) methods for a number of basic problems such as graph signal filtering or sampling \cite{shuman2013emergin,ortega2018graph}. Most of these approaches start by selecting a graph operator, typically the graph Laplacian or the adjacency matrix, and using the eigenvectors of this operator as elementary graph signal frequencies. In addition to the Laplacian and adjacency matrices, which were already studied in the context of spectral graph theory \cite{chung1996spectral}, some new graph operators have been proposed recently. Examples include operators designed to address graph signal stationarity \cite{girault2015stationary} or graph irregularity \cite{girault2018irregularity}, or leading to orthogonal basis for directed graphs \cite{shafipour2018directed}.

Letting the variable $\lambda$ denote the graph frequency. A given graph with $n$ nodes will have $n$, not necessarily distinct, discrete frequencies $(\lambda_0, \lambda_1, \ldots, \lambda_{n-1})$, each corresponding to an eigenvalue of its corresponding graph Laplacian, assuming a GFT based on the graph Laplacian.  Then, $\Um^\top$, the transpose matrix of orthogonal eigenvectors\footnote{Unless otherwise stated we will only consider undirected graphs here.}, is usually called the graph Fourier transform (GFT) \cite{shuman2013emergin,ortega2018graph}. 

The key observation in our work is that once an operator has been defined, \textit{e.g.}, the combinatorial graph Laplacian for undirected graphs, much of the recent literature has focused on developing methods that make use of that operator and are {\em generic}, in the sense of being suitable for any graph within a certain class (\textit{e.g.}, any undirected graph). 
Thus, basic GSP techniques are often designed so that exact knowledge of the distribution of eigenvalues and exact values of the eigenvectors are not required. For example, a popular design of wavelet transforms for graph signals \cite{hammond2011wavelets} is based on defining a bandpass filter $g(\lambda)$. Exact knowledge of the eigenstructure of the graph is not required as long as $g(\lambda)$ can be approximated by a polynomial. As another example, most methods proposed to date for graph signal sampling (\textit{e.g.}, \cite{chen2015discrete,anis2016efficient}) set up the problem by selecting a sampling rate (a fraction of graph nodes to be observed) and then optimizing node selection. These decisions are made under the assumption that graph signals will be smooth in some sense, \textit{e.g.}, bandlimited. The choice of sampling rate a priori does not require knowledge of the eigenstructure of the graph. While this leads to a mathematically meaningful definition of which signals can be reconstructed (\textit{e.g.,} bandlimited signals) other characteristics of these signals (such as their localization) are often unclear. 

Clearly, it is desirable to design algorithms, such as those discussed above, that can be applied to generic graphs. However, there is an inherent risk in not having sufficient knowledge about the frequency characteristics of the graph. In particular, working with signal models that can be interpreted and related to actual observed behavior becomes essential when considering specific application domains. In this paper we argue that current methods fall short in this regard and that not enough attention has been devoted to developing a more specific understanding of graph signal frequencies. For example, a given $g(\lambda)$ may have a pass band containing a relatively small number of discrete frequencies. Similarly, a given sampling rate may partition the range of frequencies in such a way that some subsets of nodes may be ``over-represented'' due to the localization properties of graph frequencies. These and similar issues arise due to the irregular nature of graph topologies, which in turn leads to irregular distribution of discrete frequencies on the real line, and localization (and even compact support) of some of the elementary frequencies. This is in contrast with elementary frequencies in conventional signal processing, which are evenly spaced and are not localized in time (or space). 

Our main goals in this paper are twofold. First, we introduce a series of novel graph frequency visualization tools, which are also being made available online\footnote{\url{https://github.com/STAC-USC/GraSP/}}. For relatively small size graphs 
these tools make it possible, for the first time, to observe specific characteristics of all graph frequencies, including their spread and vertex domain localization properties. While our proposed tools require computing the GFT, which may not be practical in some cases, and visualization may be difficult for large graphs, these ideas can be the basis to develop suitable techniques for larger graphs. 
Second, by presenting several examples of GFTs plotted using our tools, we show some key differences between GFTs that our visualization techniques can highlight, including localization of GFT eigenvectors, sampling, and localization of SGWT atoms, as a few examples.

Our work was initially inspired by techniques used to represent eigenstates in dynamical systems (\textit{e.g.}, \cite{trefethen2017exploring}). While the importance of eigenvector (graph frequency) structure in graph signal processing is undeniable, surprisingly little effort has been devoted to visualization. One exception is \cite{saito2018can}, which uses visualization examples to make the case that ordering frequencies on the eigenvalues of the Laplacian may not be adequate. Our approach also suggests that a better understanding of the implications of such an ordering is required. Unlike \cite{saito2018can} our focus is primarily visualization, with the goal of providing insights about existing methods. 

The rest of this paper is organized as follows. In Section \ref{sec:motivation} we introduce basic notation and motivate the need for visualization tools. 
In Section \ref{sec:tools}, we present our proposed visualization tools and provide simple examples comparing results for different graph operators. 
In Section \ref{sec:experiments}, we show experimental results illustrating the irregular behavior of frequencies in an example graph. These results suggest that a more detailed understanding of graph frequencies is important, before GSP methods can be applied in specific applications. Finally we conclude this paper in Section \ref{sec:conclusion}.

\section{Motivation}
\label{sec:motivation}

\subsection{Basic definitions}

We consider weighted undirected graphs $\Gc=(\Vc,\Ec,\Wm)$, where $\Vc$ is the vertex set and $\Ec$ is the edge set of the graph. The entry $w_{i,j}\geq 0$ in the weight matrix $\Wm$ represents the weight of edge $(i,j)\in\Ec$, and $w_{i,j}=0$ if $(i,j)\notin\Ec$. The graph Laplacian matrix is defined as $\Lm=\Dm-\Wm$, where $\Dm$ is the diagonal degree matrix with $d_{i,i}=\sum_{j=1}^n w_{i,j}$. Based on the definition of Laplacian matrix, for a given graph signal $\xv\in\mathbb{R}^n$, the \emph{Laplacian quadratic form}
\begin{equation}
\label{eq:lqf}
    \xv^\top\Lm\xv = \sum_{(i,j)\in\Ec} w_{i,j}(x_i-x_j)^2 
\end{equation}
measures the variation of $\xv$ on the graph. 

The graph Fourier transform (GFT) is an important tool in graph signal processing \cite{shuman2013emergin,sandryhaila2013discrete,ortega2018graph}. One possible definition of the GFT would be to choose $\Um$ the matrix of eigenvectors of the graph Laplacian: $\Lm=\Um\Lambdam\Um^\top$. Based on this definition, the GFT basis functions $\uv_0,\dots,\uv_{n-1}$ are mutually orthogonal unit norm vectors that correspond to the smallest to the largest variations on the graph:
\[
  \uv_0 = \underset{\|\fv\|=1}{\text{argmin}} \quad \fv^\top\Lm\fv, \quad
  \uv_{k} = \underset{\fv\perp \uv_0,\dots,\uv_{k-1},\|\fv\|=1} {\text{argmin}} \quad \fv^\top\Lm\fv.
\]
Each of these basis vectors corresponds to a graph frequency, \textit{i.e}, the corresponding eigenvalue $\lambda_0, \lambda_1, \ldots, \lambda_{n-1}$, where $\lambda_0=0$ corresponds to the constant eigenvector, $\uv_0 = n^{-1/2}{\bf 1}$, similar to the DC frequency for conventional signal processing. 

\subsection{Filtering}

A significant amount of research effort has been devoted to developing transformations for graph signals. As a representative example, consider  spectral graph wavelet transforms (SGWTs) \cite{hammond2011wavelets}, one of the earliest and most widely used approaches. This approach is based on designing a bandpass prototype $g(\lambda)$ (with $g(0)=0$) and a scaling function $h(\lambda)$ (with $h(0)\neq 0$). Multiple scaled versions of $g(\lambda)$, $g_t(\lambda)= g(t\lambda)$ combined with $h(\lambda)$ can be shown to form a frame, whose multi-resolution properties mimic those of wavelet transforms for conventional signals. SGWTs can be implemented in the frequency domain, by computing $\Tilde{\xv} = \Um^\top\xv$, the GFT of a signal $\xv$, and then scaling each of the frequencies by the corresponding filter weights $f(\lambda_i)$ before computing the inverse GFT. In order to avoid computing the GFT, which could be costly for large graphs, it is possible to compute the SGWT in the vertex domain by using filters $g(\lambda)$ that can be approximated by polynomials of $\lambda$. If the filters are polynomial the transformation can be applied directly in the vertex domain. 

Thus, assuming a polynomial design, one can construct a complete SGWT without having any information about the eigenvalues and eigenvectors of the graph. While not having to compute the GFT is compelling in terms of complexity, we will show that this could lead to undesirable results. For example, assume that for a specific graph and graph signal we observe that a certain band, corresponding to a scale $t$, $g_t(\lambda)$ has low energy. This could be due to the signal not having energy in those specific frequencies, or it may be due to that particular passband not containing many discrete frequencies to begin with. 

\subsection{Sampling}

As a second example, consider graph signal sampling \cite{chen2015discrete,anis2016efficient}. Some approaches require computing the GFT  (\textit{e.g.}, \cite{chen2015discrete}), while for others (\textit{e.g.}, \cite{anis2016efficient}) this is not needed. In either case, we face the question of deciding what the sampling rate should be, \textit{i.e.}, if the graph has $n$ nodes we need to choose $k < n$, the number of nodes to be sampled. 
This question is more straightforward for conventional signal sampling. $k$ allows us to quantify the frequency content of signals of interest. Since frequencies are equally spaced and all basis vectors are global (\textit{e.g.}, the DFT), it is easy to interpret the effect of increasing the number of samples. In particular, since frequencies are equally spaced, increasing the number of samples by one always increases the maximum signal frequency by the same amount (\textit{i.e.,} $1/N$-th of the maximum frequency, if the signal has length $N$).  

This is not as straightforward for graph signals. Given that frequencies are no longer evenly spaced, increasing the size of the sampling set by $1$ will increase the maximum signal frequency of a signal that can be reconstructed, but this increase will not be the same for each additional sampled node. 
As an example, consider two cases where $k_1$ and $k_2$ samples, respectively, have already been selected for sampling and we would like to choose an additional one. Then, the respective increases in the size of the signal frequency band would respectively be $\lambda_{k_1+1}-\lambda_{k_1}$ and $\lambda_{k_2+1}-\lambda_{k_2}$, which are not guaranteed to be equal. In fact, in some cases, if an eigenvalue has multiplicity greater than one, say $\lambda_{k_1+1}=\lambda_{k_1}$ there will be no increase in the size of the signal band as the size of the sampling set is increased. Moreover, since elementary graph frequencies can be  localized in the vertex domain, the choice of sampling rate can also lead to  localized sampling sets: for example, if the graph nodes can be clustered, many nodes in one cluster could be added to the sampling set, while only few will be added from another cluster.  

\section{Proposed visualization tools}
\label{sec:tools}

\begin{figure}[tb]
    \centering
    \includegraphics[width=1\linewidth,clip,trim=80 25 15 100]{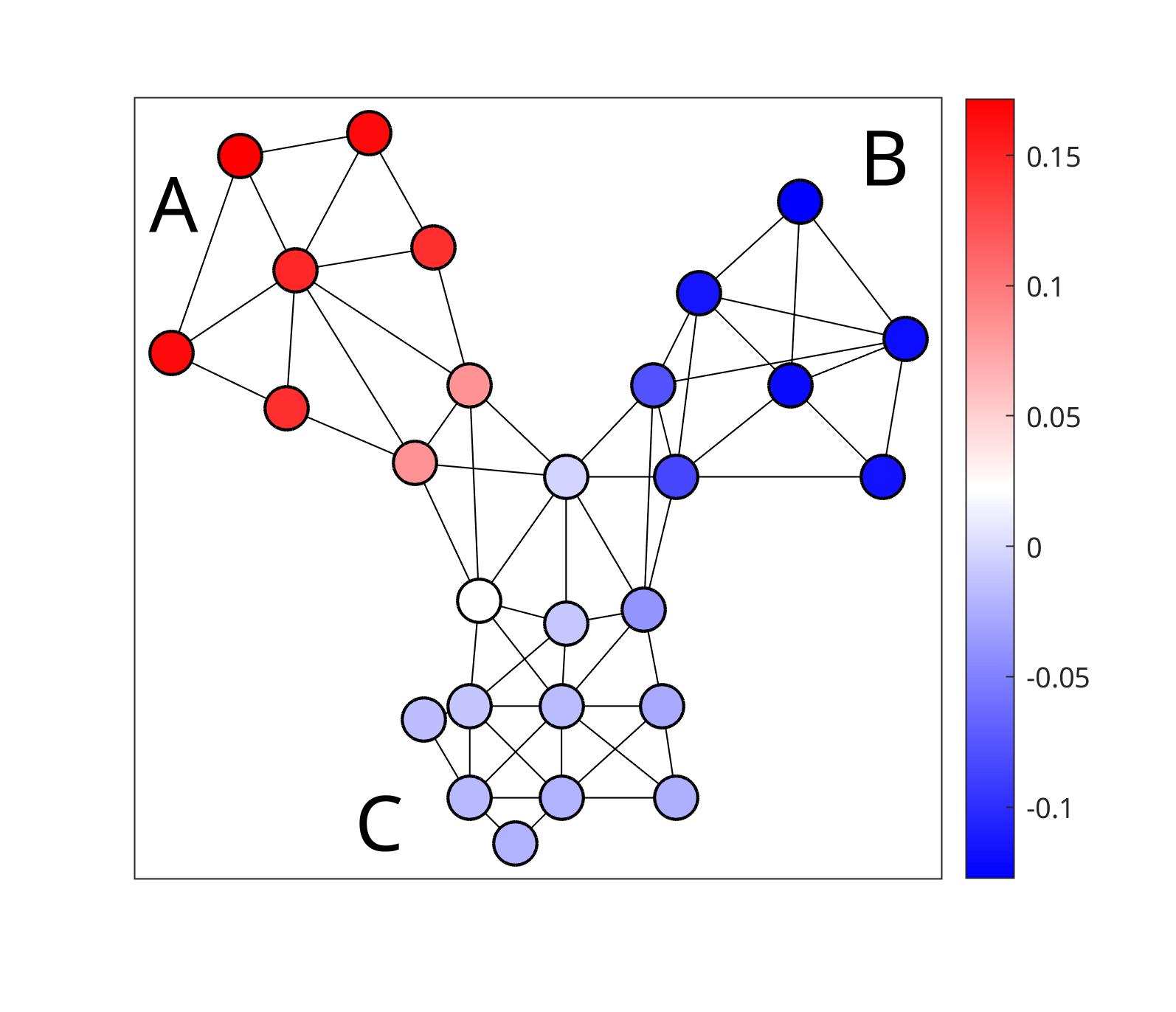}
    \caption{Unweighted toy graph example, with 1D embedding.}
    \label{fig:toy_graph:embedding}
\end{figure}

In this section, we present our proposed visualization technique, and discuss the design choices to be  made. To give a concrete example, we use the unweighted graph of \autoref{fig:toy_graph:embedding} and represent the GFT of that graph based on its combinatorial Laplacian. Focusing on the problem of displaying the GFT basis vectors,  \textit{i.e.}, the columns of $\Um$, we provide an overview of the visualization challenges and how we tackle them. Note that each of these vectors is a graph signal by itself, \textit{i.e.}, it associates a scalar value to each of the nodes.

\begin{figure}[t]
    \centering
    % \begin{subfigure}{0.49\linewidth}
    %     \centering
    %     \includegraphics[width=1\linewidth]{ToyGraphUnweighted/2D_u4_300dpi.jpg}
    %     \caption{2D Embedding}
    %     \label{fig:2Dvs1DEmbedding:2D}
    % \end{subfigure}
    % \begin{subfigure}{0.49\linewidth}
        % \centering
        \includegraphics[width=1\linewidth,clip,trim=50 0 0 80]%
            {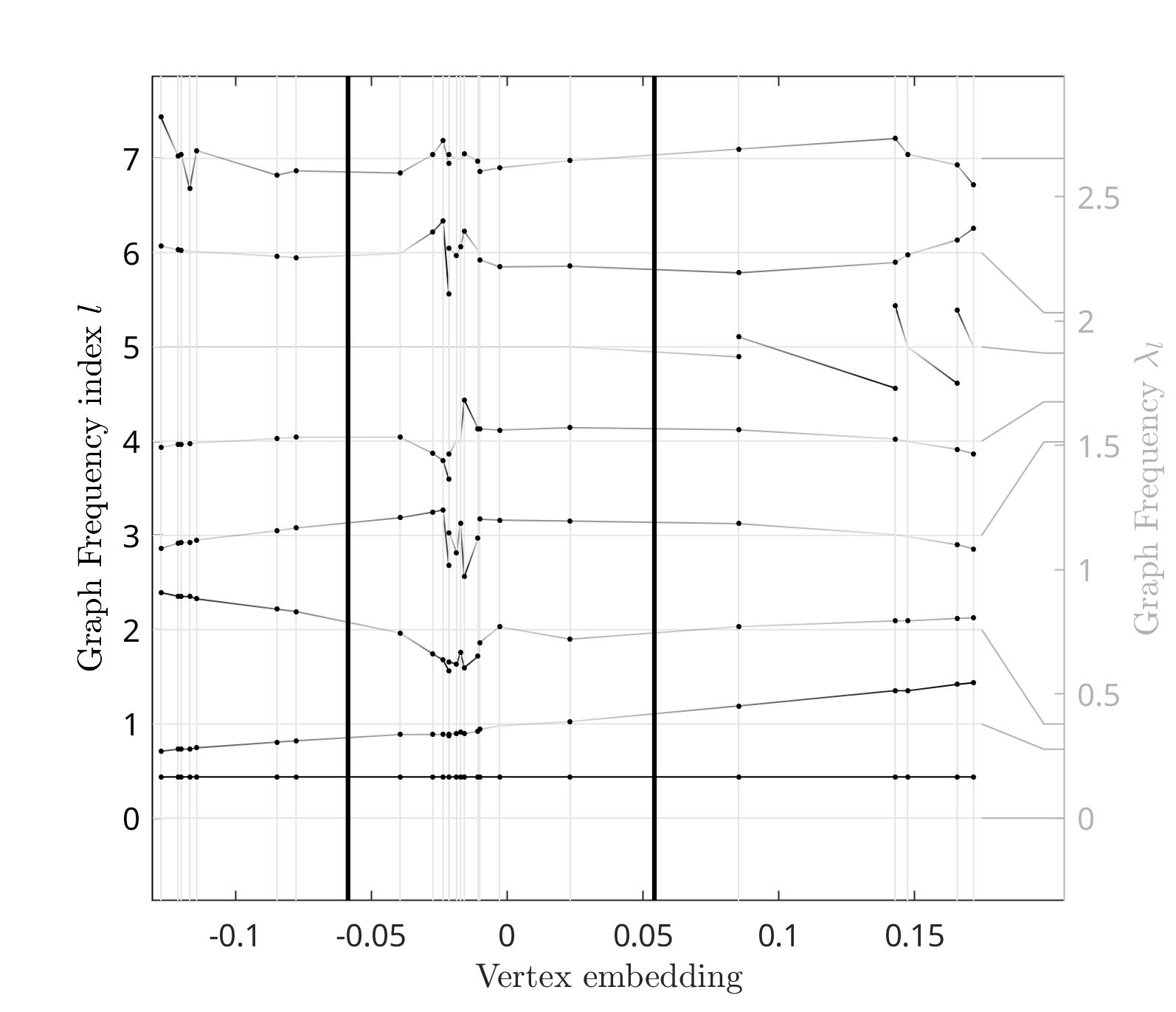}
        % \caption{1D Embedding}
        % \label{fig:2Dvs1DEmbedding:1D}
    % \end{subfigure}
    \caption{Partial combinatorial Laplacian GFT of the unweighted toy graph in \autoref{fig:toy_graph:embedding}. Only the first 8 GFT basis vectors are shown here:  $\{\uv_0,\dots,\uv_7\}$, with \lstinline{'amplitude_scale'} set to 1 (no overlap between GFT basis vectors).}
    % \label{fig:2Dvs1DEmbedding}
    \label{fig:toy_graph:partial_gft}
\end{figure}

\begin{figure}[t]
    \centering
    \includegraphics[width=1\linewidth,clip,trim=80 25 15 100]%
        {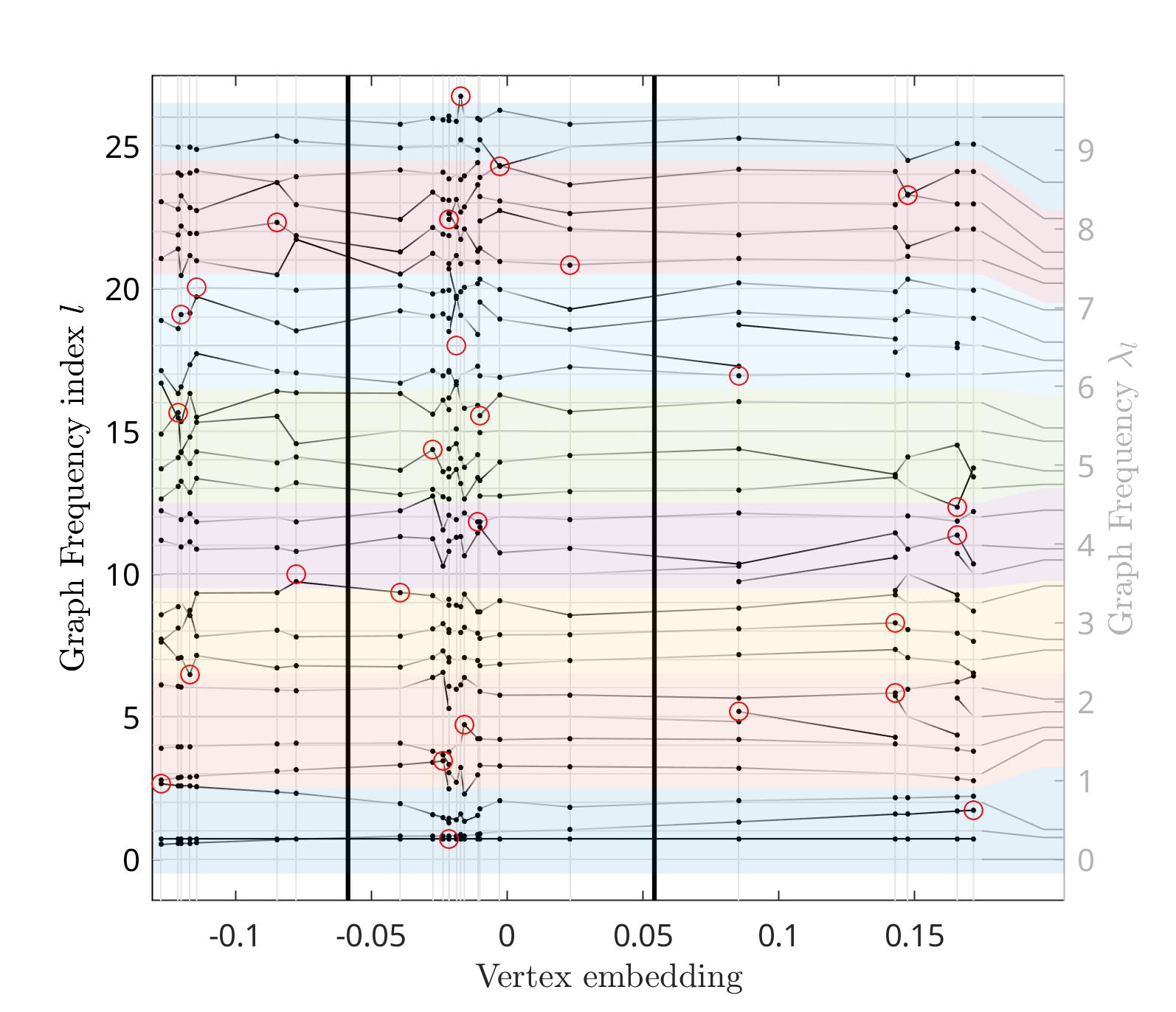}
    \vspace{0.01em}
    \caption{Combinatorial Laplacian GFT visualization using the graph of \autoref{fig:toy_graph:embedding}. The visualization includes regularly spaced spectral bands between the lowest graph frequency (0) and the maximum. The red circles correspond to a sampling using \cite{anis2016efficient} (cf. \autoref{sec:experiments:sampling} for details).}
    \label{fig:toy_graph:comb_lapl}
\end{figure}

\begin{figure*}[tb]
\begin{lstlisting}
grasp_start_opt_3rd_party('usc_graphs');
load('toy_graph.mat');
toy_graph = grasp_eigendecomposition(toy_graph);
sampling_sets = gsp_sampling(toy_graph);
clusters = [ones(12, 1) ; 2 * ones(8, 1) ; 3 * ones(7, 1)];
grasp_show_transform(gcf, toy_graph,...
    'clusters', clusters,...
    'embedding', 0,...
    'bands', max(toy_graph.eigvals) / 8 * [(0:7)@sq@ (1:8)@sq@],...
    'highlight_entries', sampling_sets);
\end{lstlisting}
    \caption{\matlab code to generate \autoref{fig:toy_graph:comb_lapl} using \grasp version 1.2.0. \texttt{embedding} is set to 0 to automatically compute the embedding from the second eigenvector of the random walk Laplacian. \texttt{gsp\_sampling} is a generic function that returns a matrix $\Sm$ such that $[\Sm]_{li}$ is one whenever $i$ is the the $l^{\text{th}}$ sampled vertex (and 0 otherwise). See \autoref{sec:experiments:sampling} for details.}
    \label{fig:toy_graph:comb_lapl_code}
\end{figure*}

\subsection{Visualization challenges and design choices}

\paragraph*{Embedding}
Note that there is no ``natural'' way to plot a graph signal. Some graphs can be plotted in 2D space, \textit{i.e.}, each node is given horizontal and vertical coordinates, and the graph signal can be represented as a node attribute (see \autoref{fig:toy_graph:embedding}). However, nodes are not necessarily points in a metric space, and if they are, the dimension of the space could be high. In order to achieve a compact representation, here we choose to map all graph nodes onto the real line via an \emph{embedding}. Clearly, information will be lost in the embedding, but since our goal is to compare the GFT basis vectors, using the same embedding for all of them will provide an easy and consistent way to visualize them. Note that, although the embedding may not be that robust in terms ordering individual nodes, it will generally preserve clusters. It is the global structure associated to clusters that matters most to understand graph frequencies and the GFT.

\paragraph*{Frequency spacing}
Graph frequencies (the $\lambda_k$'s) are not equally spaced. We decided to display each of the GFT basis vectors as a signal on the real line, and to stack them all vertically. These are the horizontal signals plotted on \autoref{fig:toy_graph:partial_gft}. When doing so one option is to make the vertical coordinate of each component $\uv_k$ a function of the corresponding $\lambda_k$. This proved not to be practical for graphs with very irregular frequency spacing, or when there are numerous eigenvalues with multiplicity greater than one, since basis vectors corresponding to similar (or identical) frequencies would be hard to distinguish (see for example \autoref{fig:gft_weighted_toy_aux:comb_yfreq}). Instead, we chose to use the index of the component ($k$ for a given $\uv_k$) to space them (\textit{i.e.}, they are equispaced). Then, in order to convey the irregular spacing of eigenvalues we plot them along the vertical right axis in their original spacing and plot a line linking each component to the corresponding graph frequency. Notice that in the case of a more general graph signal linear transform, one has to associate a graph frequency to each of the transform components. We will give an example of such an association in the case of the SGWT in \autoref{sec:experiments:sgwt}.

\paragraph*{Magnitude} 
We had to decide on how to convey the magnitude of the value associated to each node for a given GFT basis vector. That is, denoting $\uv_k$ the k-th eigenvector, we are interested in plotting its $i^{\text{th}}$ entry $[\uv_k]_i$ in the position of the $i^{\text{th}}$ node in our embedding. To do this, we first considered plotting this information as a signal with each sample connected to the next using a black line. 
Unfortunately, these black lines were too distracting, since many of the magnitudes are actually very close to zero (see for example $\uv_5$ on \autoref{fig:toy_graph:partial_gft}). To better see the support without the distracting influence of nodes with zero magnitude, 
we use a shade of gray that corresponds to the magnitude: highest magnitude is black, zero is a light gray, and lines between two consecutive nodes in the embedding have a shade of gray that is interpolated between the magnitudes (see \autoref{fig:toy_graph:partial_gft} for an illustration). Moreover, since all zero or close to zero entries are difficult to identify, we implemented two additional features to better highlight the support of the signals. The most noticeable one in \autoref{fig:toy_graph:partial_gft} are the dots. These are added each time the magnitude is greater than some threshold parameter (see \autoref{sec:tools:implementation}), hence showing the support of the signals. The second feature is the vertical light gray bars: these show the embedding. In addition, these vertical bars can be used to spot the entries that are not in the support of a GFT basis vector: a lack of dot around the intersection the vertical bar and horizontal line of a GFT basis vector corresponds to approximately zero magnitude.

\paragraph*{Additional information for visualization}
Along with the basic information described in the previous paragraph, we decided to add visual elements to make explicit the presence of structure in the graph. In particular, we included an option to plot cluster boundaries, which are shown as thick black vertical bars in \autoref{fig:toy_graph:partial_gft}.
There are additional visualization features that will be discussed in the following sections. Two of these are of particular interest and are shown in \autoref{fig:toy_graph:comb_lapl}: spectral bands represented by colored regions of the plot, and highlighted entries using red circles (useful in the context of sampling). We next illustrate these choices with an example, 
followed by details about our implementation in Section~\ref{sec:tools:implementation}.

\subsection{Illustration on a Toy Graph}

\def\showtransform{\lstinline{grasp_show_transform}\xspace}

The function \showtransform of the \grasp toolbox \cite{Girault.ICASSPDEMO.2017} implements our proposed visualization technique\footnote{Available since version 1.2.0.}. In \autoref{fig:toy_graph:comb_lapl} we show the visualization obtained for the graph Fourier transform based on the combinatorial Laplacian of the unweighted graph shown in \autoref{fig:toy_graph:embedding}, and using the code in \autoref{fig:toy_graph:comb_lapl_code}. Additionally, we show in \autoref{fig:toy_graph:partial_gft} a subset of the GFT basis vectors to better illustrate key properties of the embedding and representation of the magnitudes.

Several features are of importance. First and foremost, we can see the stacked GFT basis vectors, appearing as irregularly sampled times series thanks to the 1D embedding of nodes. Note the light gray horizontal lines that depict for each of the basis vectors the value 0, and the light gray vertical lines that identify each vertex embedding. These pseudo-time series continue then to the right of the figure until they are mapped on the right axis to the associated graph frequency. We notice then the importance of this mapping due to the irregular spacing of graph frequencies.

In \autoref{fig:toy_graph:comb_lapl}, the frequencies are then grouped into equal size bands using colored backgrounds, leading to colored groups of GFT basis vectors. This shows an effect of the irregular nature of graph frequencies: even though the bands are of equal size, they do not correspond the same number of GFT basis vector.

Next, the approximate support of the GFT basis vector is depicted using black dots: each time an entry has a large enough magnitude a dot is added. Note that some of the GFT basis vector have a very small support (\textit{e.g.} $\uv_5$). This is one property that makes GFT basis vectors very different from Fourier modes in conventional signal processing. Making it possible to visualize this behavior is one of the main motivations for our work. 

Finally, the clusters are separated by thick black vertical lines, and some of the entries of the GFT matrix are highlighted using red circles in \autoref{fig:toy_graph:comb_lapl}. These highlighted entries can be used to represent a specific sampling algorithm. Assuming that a sampling algorithm selects one node to sample at each iteration, we represent the first chosen node by highlighting the corresponding entry in the first GFT basis vector, the second selected sample is shown in the second GFT basis vector, and so on. Each additional node sampled allows for an increase in bandwidth of the graph signal we can reconstruct. 

\subsection{Parameters of the \grasp implementation}
\label{sec:tools:implementation}

The \grasp toolbox implements our proposed visualization technique in the function \showtransform. This section provides a description of visualization parameters 
and serves as a reference of the implementation in version 1.2.0 of \grasp.

\paragraph*{Embedding}
By default, \showtransform uses the second eigenvector of the random walk Laplacian, a typical choice in the literature \cite{Belkin.NEURCOMP.2003}, to generate a 1D embedding of the vertices. 
To override that choice, the parameter \lstinline{'embedding'} can be used, and can be set to either
\begin{inlinelist}
    \item \lstinline{0} (default) for an embedding equal to the second eigenvector of the random walk Laplacian
    \item \lstinline{1} for a regular (equispaced) embedding based on the second eigenvector of the random walk Laplacian
    \item some $n$ dimensional vector for an arbitrary embedding
\end{inlinelist}.

\paragraph*{Clusters}
The parameter \lstinline{'clusters'} can be set as a vector giving the cluster id of each vertex. In that case, the resulting figure will show thick vertical bars (see \autoref{fig:toy_graph:comb_lapl}) between clusters. Alternatively, a single integer value $k$ will trigger the computation of $k$ clusters using an the embedding in $k$D based on the random walk Laplacian eigenvectors. 

\paragraph*{Magnitude Normalization}
If the magnitude of the GFT basis vectors is not constrained, the visualization can quickly become hard to read with too much overlap between stacked GFT basis vectors (magnitudes too large), or details too small to read (magnitudes too small). To avoid this problem, we implemented three normalization schemes for the GFT basis vectors to plot. The parameter \lstinline{'amplitude_normalization'}\footnote{\emph{Amplitude} refers here to the maximum magnitude over vertices of a mode.} takes three possible values: 
\begin{inlinelist}
    \item \lstinline{'l2'} to have basis vectors of unit norm, $\ell_2(\uv_k)=1, \forall k$ 
    \item \lstinline{'max_abs'} (default) to have basis vectors of maximum unit amplitude, $ \ell_\infty(\uv_k)=1, \forall k$ 
    \item \lstinline{'overall_max_abs'} to have the maximum amplitude over all modes equal to 1, $\max_k \ell_\infty(\uv_k)=1$, (same normalization factor for all modes) 
\end{inlinelist}.
This constrains the magnitude to a predefined range, and the various options above control the relative amplitude normalization between GFT basis vectors. Once the amplitudes are normalized, the value of the parameter \lstinline{'amplitude_scale'} controls the overlap between the GFT basis vectors: a value 1 means no overlap, and 1.5 a 50\% overlap (anything smaller 1 corresponds then to a gap between GFT basis vectors).

\paragraph*{Support Visualization} 
Often the support of a GFT basis vector spans the entire vertex set. However, in some cases the magnitude at a specific node may be so small that considering the node to be in the support may be undesirable. To account for this, we propose the parameter \lstinline{'epsilon_support'} as a threshold on the magnitude to decide whether a given vertex is in the approximate support of the mode. This is based on the values after normalization (see \lstinline{'amplitude_normalization'} above), and by default the threshold is 0.05 (\textit{e.g.} 5\% of the the maximum magnitude of the mode for \lstinline{'max_abs'}). Moreover, the parameter \lstinline{'support_scatter_size'} controls the size of the dots shown for the vertices in the support of the modes (default value: 36).

\paragraph*{Bands} A bandpass filter is a filter that attenuates the graph signal spectral components outside of its band. Given such a filter, in order to visualize its effect on the GFT basis vectors, we added the possibility to highlight a band, \textit{i.e.}, a given interval of frequencies. The parameter \lstinline{'bands'} can be set to a matrix with 2 columns, one for the beginning of the band, and one with the end of the band, for as many bands as rows in the matrix. Each band is then colored similarly to what we observe on \autoref{fig:toy_graph:comb_lapl} (shown here with 8 bands of equal size and spanning the entire spectrum). Additionally, colors can be set using the parameter \lstinline{'bands_colors'} and providing a matrix with 3 columns (RGB channels) and as many rows as there are bands to plot.

\paragraph*{Vertical Alignment of GFT Basis Vectors} As described earlier, we decided to represent each of the GFT basis vectors equispaced along the vertical direction. However, we also implemented the scheme where each mode is aligned vertically with its graph frequency, for reference. This is set using the parameter \lstinline{'graph_signal_y_scheme'} set to \lstinline{'freq'} (this parameter default value is \lstinline{'regular'} for equispaced modes). This can serve as a reference and may also be useful in instances where information about specific basis vectors is not as important and thus overlap could be acceptable.  

\paragraph*{Alternative Transform}
\showtransform displays by default the GFT of the graph given as argument. To visualize another linear transform given by a matrix $\Hm$ of size $M\times N$, one can set the parameter \lstinline{'transform_matrix'} to \lstinline{H}. The columns of $\Hm^*$ are then used instead of the GFT basis vectors. We will call these columns the \emph{atoms} of the linear transform\footnote{Note that in case of using an inner product for graph signals $\Qm\neq\Id$ (\textit{i.e.}, not the dot product), then the atoms are given by the columns of the matrix $\Qm^{-1}\Hm^*$. See \cite{girault2018irregularity} for details.}. Note that $\Hm$ is effectively an \emph{analysis} transform. For example, in the case of the SGWT discussed in \autoref{sec:experiments:sgwt}, the transform is defined as the orthogonal projection on the set of localized and scaled wavelet atoms. This defines a transform matrix that given an input graph signal $\xv$, outputs the set of wavelet coefficients $\Hm\xv$.

\paragraph*{Custom Frequencies} When providing a custom graph transform matrix using \lstinline{'transform_matrix'}, the atoms of the transform are not naturally associated with a set of graph frequencies, since the atoms of the transform may not overlap with the GFT basis vectors. For example, the SGWT uses bandpass filters, and we may be interested in mapping each SGWT atom to some graph frequency that depends on the band of the filter. For this purpose, the parameter \lstinline{'graph_frequencies'} can be set to a vector of frequencies, with as many entries as there are distinct frequencies. In \autoref{sec:experiments:sgwt}, we use the frequency of maximum response of the bandpass filter to map each atom to a graph frequency.

\paragraph*{Sampling and Highlighted Entries}
In the context of sampling on graphs, one can be interested in understanding the relation between the spectrum of a graph and the nodes sampled (see \autoref{sec:experiments:sampling} for details). To achieve this we created an option to highlight some of the entries of the transform matrix. The parameter \lstinline{'highlight_entries'} can be set to a matrix of 0/1 values of the same size as the transform matrix, where the $(l,i)$ entry is one when the corresponding entry in the transform matrix needs to be highlighted (red circles on \autoref{fig:toy_graph:comb_lapl}). In the case of the GFT, this would highlight the entry $[\uv_l]_i$.

\paragraph*{Verbosity}
Similar to several other functions in the \grasp toolbox, verbosity of the function \lstinline{grasp_show_transform} can be set using the parameter \lstinline{'verbose'}. By default it is set to \lstinline{true} as it prints important information.

\paragraph*{Future Features} \grasp being an actively developed toolbox, its features can change overtime. Documentation is however provided for each function, and we encourage the interested reader to look at the source code for documentation of the various parameters of any function in the toolbox, including \lstinline{grasp_show_transform}.

\section{Experimental Results}
\label{sec:experiments}

\begin{figure*}[tb]
    \centering
    \begin{subfigure}{.49\linewidth}
        \includegraphics[width=\linewidth]{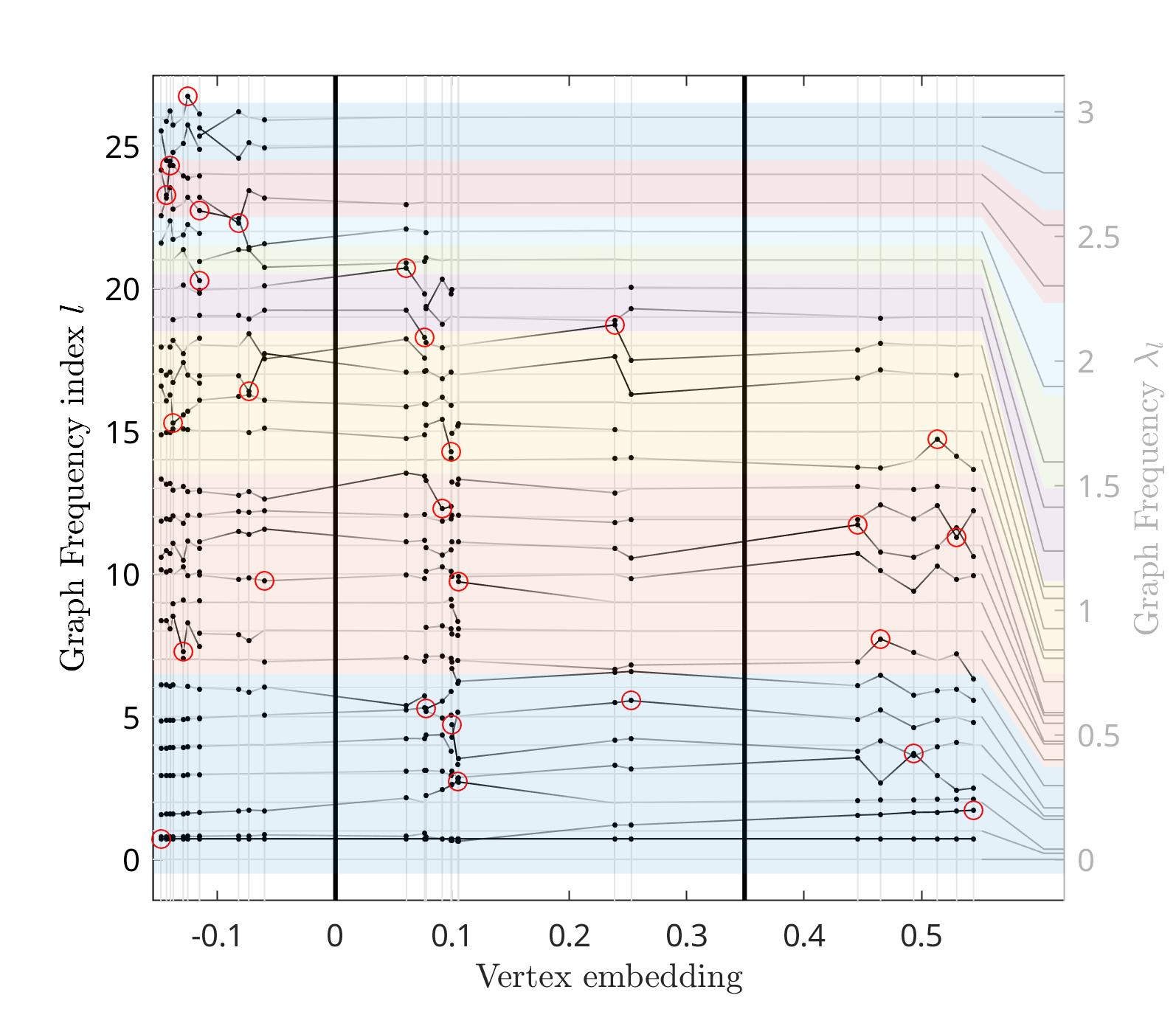}
        \caption{Combinatorial Laplacian GFT}
        \label{fig:gft_weighted_toy:comb_lapl}
    \end{subfigure}%
    \begin{subfigure}{.49\linewidth}
        \includegraphics[width=\linewidth]{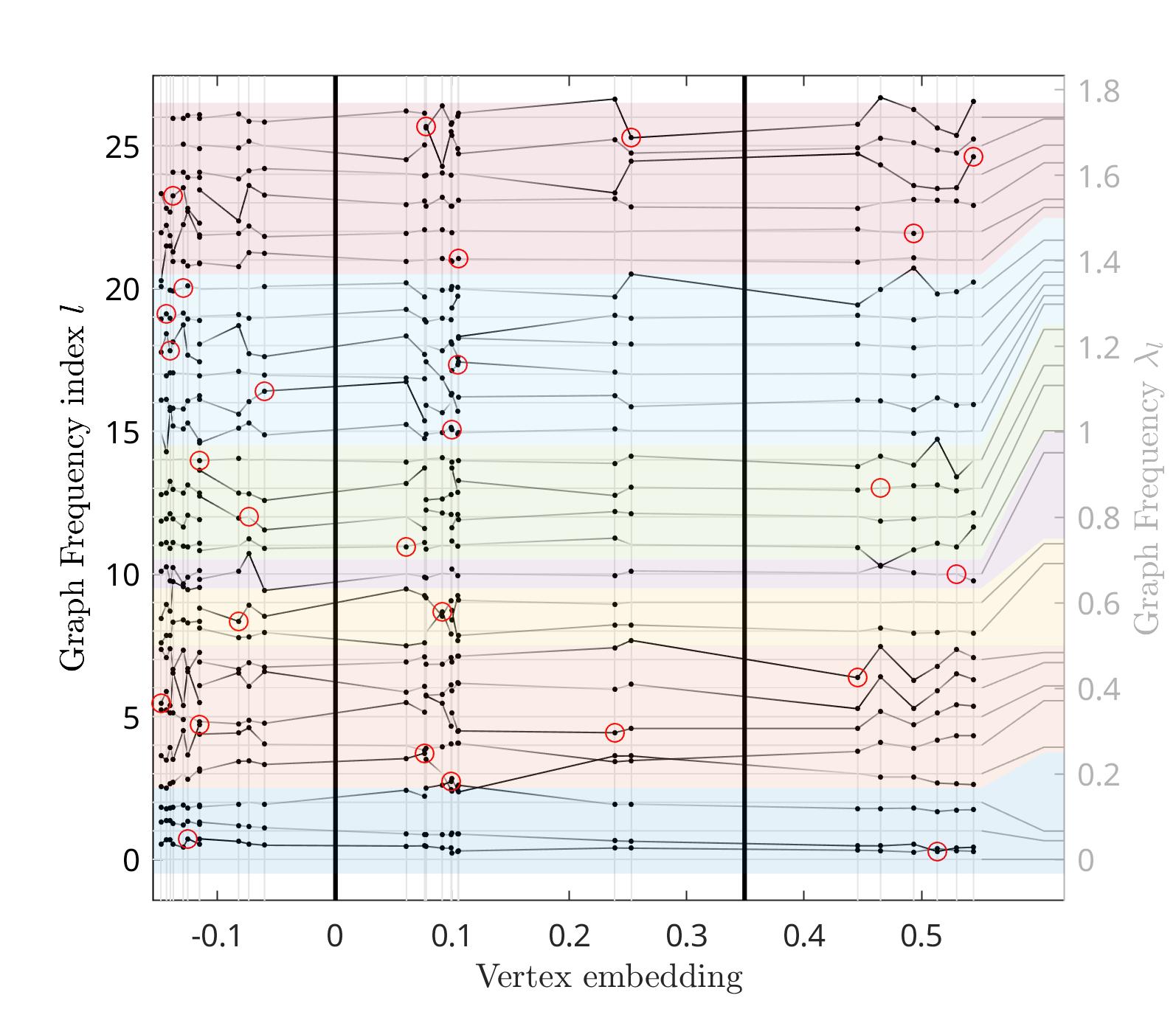}
        \caption{Normalized Laplacian GFT}
        \label{fig:gft_weighted_toy:norm_lapl}
    \end{subfigure}
    \caption{GFT visualization for \subref{fig:gft_weighted_toy:comb_lapl} the combinatorial Laplacian approach and \subref{fig:gft_weighted_toy:norm_lapl} the normalized Laplacian approach.}
    \label{fig:gft_weighted_toy}
\end{figure*}

In this section, we use again our toy graph, but this time we are giving weights to the edges. More precisely, this graph consists of nodes in 2D and we choose the edge weights using a Gaussian kernel:
\[
    \Wm_{ij}=\exp\left(\frac{\dist(i,j)^2}{2\sigma^2}\right)
    \text{,}
\]
where we chose $\sigma=1.5$ and $\dist(i,j)$ is the Euclidean distance between $i$ and $j$. This setting allows us to highlight important properties we wish to study.

\begin{figure*}[tb]
    \centering
    \begin{subfigure}{.49\linewidth}
        \includegraphics[width=\linewidth]{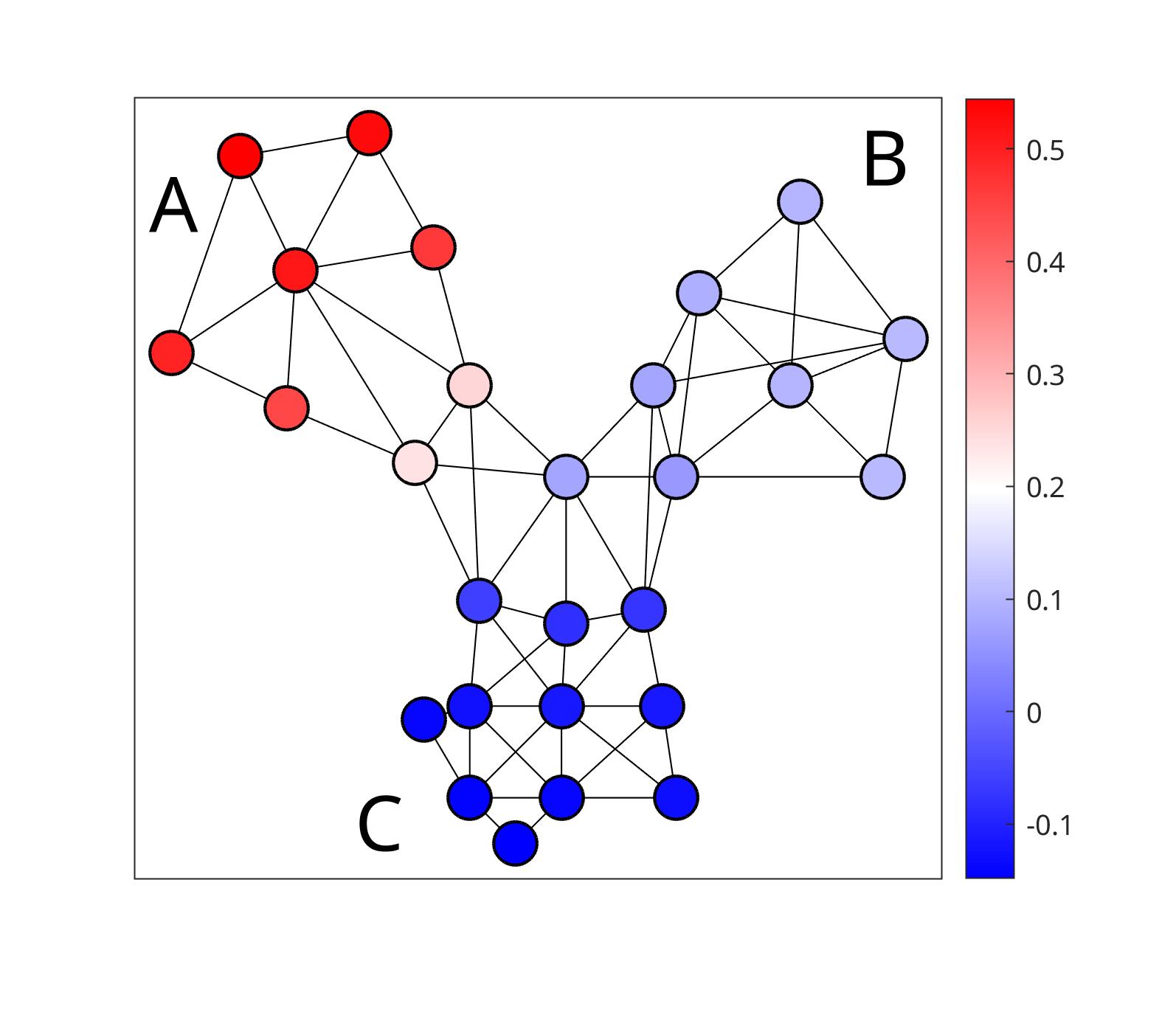}
        \caption{Embedding}
        \label{fig:gft_weighted_toy_aux:embedding}
    \end{subfigure}%
    \begin{subfigure}{.49\linewidth}
        \includegraphics[width=\linewidth]{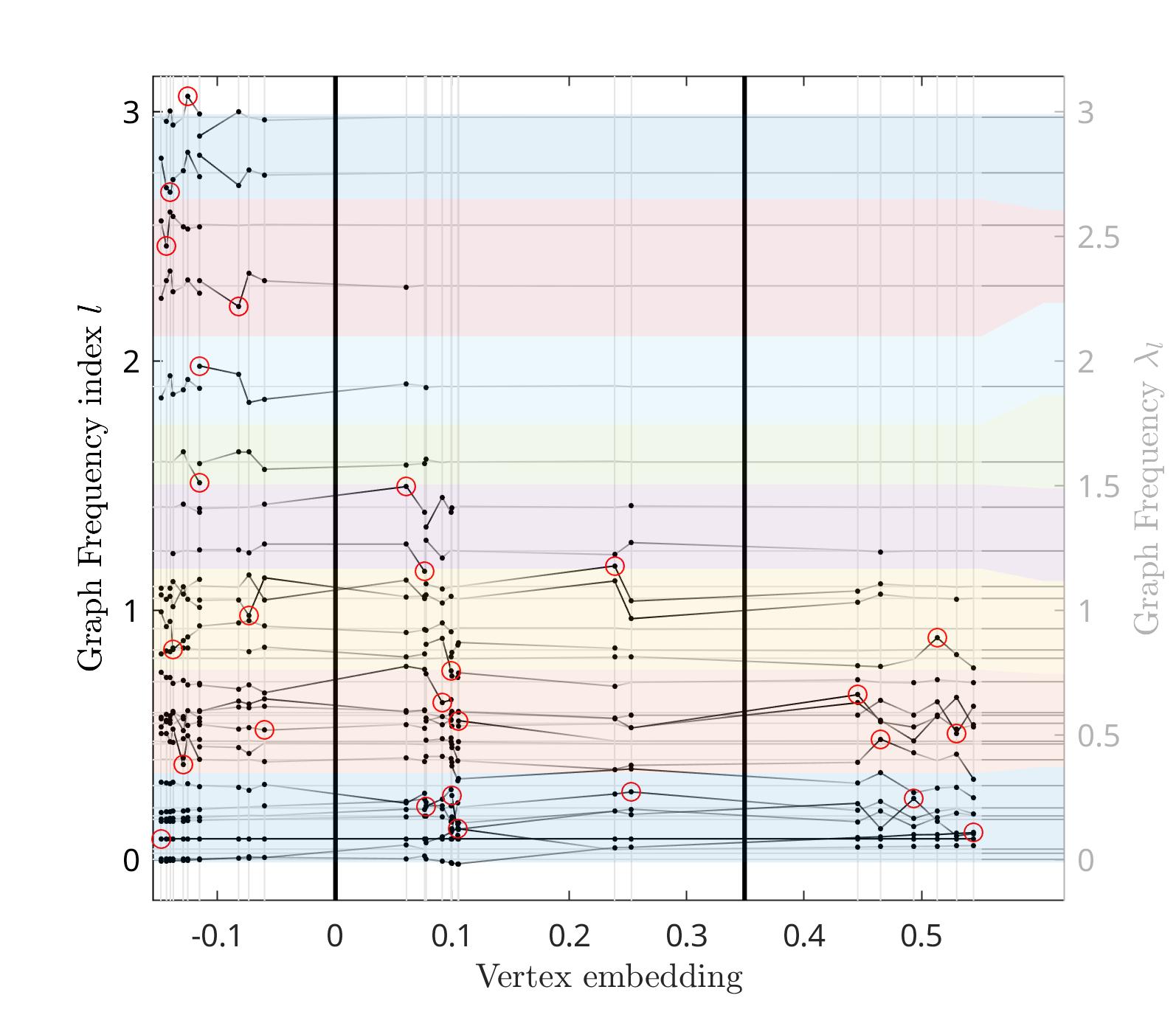}
        \caption{Combinatorial Laplacian}
        \label{fig:gft_weighted_toy_aux:comb_yfreq}
    \end{subfigure}
    \caption{\subref{fig:gft_weighted_toy_aux:embedding} Embedding used in \autoref{fig:gft_weighted_toy} and in \subref{fig:gft_weighted_toy_aux:comb_yfreq} which is the same as \autoref{fig:gft_weighted_toy:comb_lapl} with GFT basis vectors aligned vertically with their graph frequency instead of equispaced GFT basis vectors. Notice how cluster A is mapped to the left section of \subref{fig:gft_weighted_toy_aux:comb_yfreq} and \autoref{fig:gft_weighted_toy}, cluster B to the right section, and cluster C to the middle section.}
    \label{fig:gft_weighted_toy_aux}
\end{figure*}

Before delving into the visualization, notice how the nodes are grouped into three natural clusters in \autoref{fig:gft_weighted_toy_aux:embedding}: the cluster A with an embedding with large values is mapped to right hand side of \autoref{fig:gft_weighted_toy}, the cluster C with an embedding with the smallest values is mapped to the left hand side of \autoref{fig:gft_weighted_toy}, and finally the cluster B is mapped to middle section of \autoref{fig:gft_weighted_toy}.

\subsection{GFT}
\label{sec:experiments:gft}

For this first experiment, we want to highlight a key difference between the classical Fourier transform and the GFT: the GFT basis vectors can be exactly (or approximately) compactly supported. Indeed, as the literature shows \cite{Agaskar.IEEEINFTH.2013,Pasdeloup.EUSIPCO.2015,Pasdeloup.ICASSP.2016,Tsitsvero.TSP.2016,Perraudin.APSIPATSIP.2018}, uncertainty principles for GFT basis vectors allow them to be localized in the vertex domain.

The GFT based on the combinatorial Laplacian for the weighted toy graph shows such a property. In \autoref{fig:gft_weighted_toy:comb_lapl}, cluster A is not in the support of GFT basis vectors associated to the highest frequencies in the spectrum. In other words, with such a GFT, an ideal high pass filter that filters out any frequency below 1.5 (purple band and below) is actually removing all energy that the signal has on any of the cluster A nodes. Therefore, such a high pass filter is mostly useful to describe high variations on cluster C (a couple of nodes in cluster B have some energy in the support).

This example also shows that a signal that has energy that is completely localized in cluster A cannot exhibit frequencies greater than a certain threshold. This does not mean that the signal is bandlimited in the conventional sense, it just means that signals with energy only in that cluster can only achieve a certain maximum frequency. Here, \autoref{fig:gft_weighted_toy:comb_lapl} shows that the spectral band of such a signal is actually smaller than that of a signal localized in cluster C.  In general, this case shows that if we have $\hat{\xv}(k)=\langle \xv,\uv_k\rangle\simeq 0$, this could correspond to two distinct cases: either $\xv$ is orthogonal to $\uv_k$, or $\xv$ is localized on some vertices that are not in the support of $\uv_k$. Strictly speaking, of course, both are instances of orthogonality, but in the latter case we get information about localization: no signal compactly supported in that region would have any energy along that basis vector. 

On the other hand, for this example, the normalized Laplacian approach in \autoref{fig:gft_weighted_toy:norm_lapl} does not show such behavior, and most of the GFT basis vectors have support across most of the nodes. Note that this simple experiment cannot be used to generalize. A more thorough study needs to be performed to draw any conclusion on the topic of combinatorial versus normalized Laplacian with respect to GFT basis vector support.

\subsection{Sampling}
\label{sec:experiments:sampling}

Our visualization technique can also be leveraged to understand sampling schemes. Typical algorithms for sampling set selection choose $k$ out of the $n$ nodes in order to optimize interpolation to the $n$ nodes from data observed on the chosen $k$. In this formulation, $k$ can be viewed as the largest frequency index for a bandlimited signal that could be reconstructed from $k$ samples. Assuming a greedy procedure such as \cite{anis2016efficient} is used, 
we can use the parameter \lstinline{'highlight_entries'} that highlights $[\uv_k]_i$ if the $(k+1)^\text{th}$ sampled node is $i$, for all $k\in\{0,\dots,n-1\}$. 

We implemented this using the sampling method in \cite{gadde2014active,anis2016efficient} and show the results in  \autoref{fig:gft_weighted_toy}.  These results show that the difference in sampling behavior depending on which Laplacian is selected. In particular, when the combinatorial Laplacian is used and the graphs have clusters with different densities, we can see that the samples are not uniformly distributed across the clusters. Indeed we can see that the nodes on the left of the GFT plot (corresponding to the denser cluster C in the graph), continue to be sampled even after the other two clusters have been completely sampled. Since edge weights are a function of distance between node, and we expect better interpolation if nodes are close, this is a sensible result: data from nodes in the dense cluster is easier to interpolate, and thus nodes in that cluster are given lower priority for sampling. 

\subsection{SGWT}
\label{sec:experiments:sgwt}

In this experiment, we are interested in studying the support of the SGWT atoms across scales and nodes in the graph. More precisely SGWT atoms are formally defined as the impulse response of bandpass filters \cite{hammond2011wavelets}:
\[
    \gv_{m,i}=\Gm_m\deltav_i
    \text{,}
\]
where $\Gm_m=g_m(\Lm)=g(t_m\Lm)$ is the matrix of a bandpass filter whose band is defined by the function $g$ and the scaling factor $t_m$. We are interested in the support of the atoms $\gv_{m,i}$, and in particular whether we can see differences in support at coarse and fine scales. 

In the classical setting, a wavelet atom is merely a scaled and translated version of a mother wavelet \cite{Mallat.BOOK.1999}. Two observations are important here:
\begin{inlinelist}
    \item the scale of the wavelet completely defines the size of the support of said wavelet
    \item the translation operator does not change the magnitude of the Fourier coefficient of the scaled wavelet
\end{inlinelist}.
In other words, the scale completely defines the support in both the direct and spectral domains. 
However, as we saw in \autoref{sec:experiments:gft} for the GFT, some spectral bands may have a localized support. As such, the impulse responses of the bandpass graph filter $\Gm_m$ have varying support, especially for the combinatorial Laplacian approach: depending on which node the atom is centered on, the support of the atom may have varying size on the graph.

This is validated with \autoref{fig:sgwt_atoms:comb_lapl}, where the sets of atoms $\{\gv_{m,i}\}_{m,i}$ for each scale are stacked, from coarsest scale on the bottom to finest scale on top. Within each scale, the ordering of the nodes is based on  the 1D embedding: if node $i$ is assigned to a real value $e_i$ in the 1D embedding, then the corresponding atom will be plotted above all nodes $j$ assigned embedding locations $e_j$ such that $e_j<e_i$. Finally, the values are normalized using 
\lstinline{'overall_max_abs'} to maintain the relative amplitude difference between atoms.

Two observations are of importance in \autoref{fig:sgwt_atoms:comb_lapl}. First, as expected, most atoms of cluster B are vanishing. This finest scale corresponds to a high pass filter where the support is limited to cluster C. Therefore, the filter impulse response on those nodes of cluster A is close to zero. This property is not shared by the nodes of the cluster C (top left corner of the figure), where many of the atoms have several nodes in their support. Given these observation, it is clear that the study of the SGWT decomposition of a graph signal needs to take into account the support of the atoms themselves: a zero in the decomposition may correspond to either the signal being orthogonal to the atom, or the atom itself being zero in the vertices where the signal is localized.

The second observation is more significant. Indeed, even in the coarsest wavelet scale (second row of atoms from the bottom in \autoref{fig:sgwt_atoms:comb_lapl}), some atoms have zero support. This is the case for most atoms centered in cluster C. This property needs some thorough investigation, however, we can already identify some explanation using the GFT. In \autoref{fig:sgwt_bands:comb_lapl}, we plotted, for each scale, the output of the corresponding bandpass filter $g_m$ applied to the GFT (each row corresponds to $\Gm_m\uv_k$ with $\uv_k$ a GFT basis vector). For cluster C, it is quite clear that most of the nodes of the cluster have almost zero magnitude on this filtered GFT basis. In other words, since we have:
\[
    \gv_{m,i} = \sum_k \langle \delta_i, g_m(\lambda_k)\uv_k\rangle
    \text{,}
\]
and $g_m(\lambda_k)\uv_k$ is vanishing on nodes $i$ in cluster C, then the $\gv_{m,i}$ are also vanishing. Therefore, the property of non consistent support for a given scale can happen for coarser and finer scales alike.

To end this section, we observe that the SGWT based on the normalized Laplacian is slightly better with more consistent support size, but it still shows some issues (see \autoref{fig:sgwt_atoms:norm_lapl}). It should be noted that spectrum adapted tight graph wavelets \cite{Shuman.TSP.2015} tackle part of the issue, where some bands defined by the SGWT may actually correspond to a very small set of graph frequencies\footnote{This issue is not arising clearly here since the graph frequency distribution is not skewed as much as with other graphs (see \autoref{fig:gft_weighted_toy:comb_lapl}).}, but it does not solve the issue of localized GFT basis vectors, which is the root cause of some of the vanishing SGWT atom.

\def\sgwtwidth{1}
\begin{figure*}
    \centering
    \begin{subfigure}{0.49\linewidth}
        \centering
        \includegraphics[width=\sgwtwidth\linewidth,clip,trim=0 300 0 300]%
            {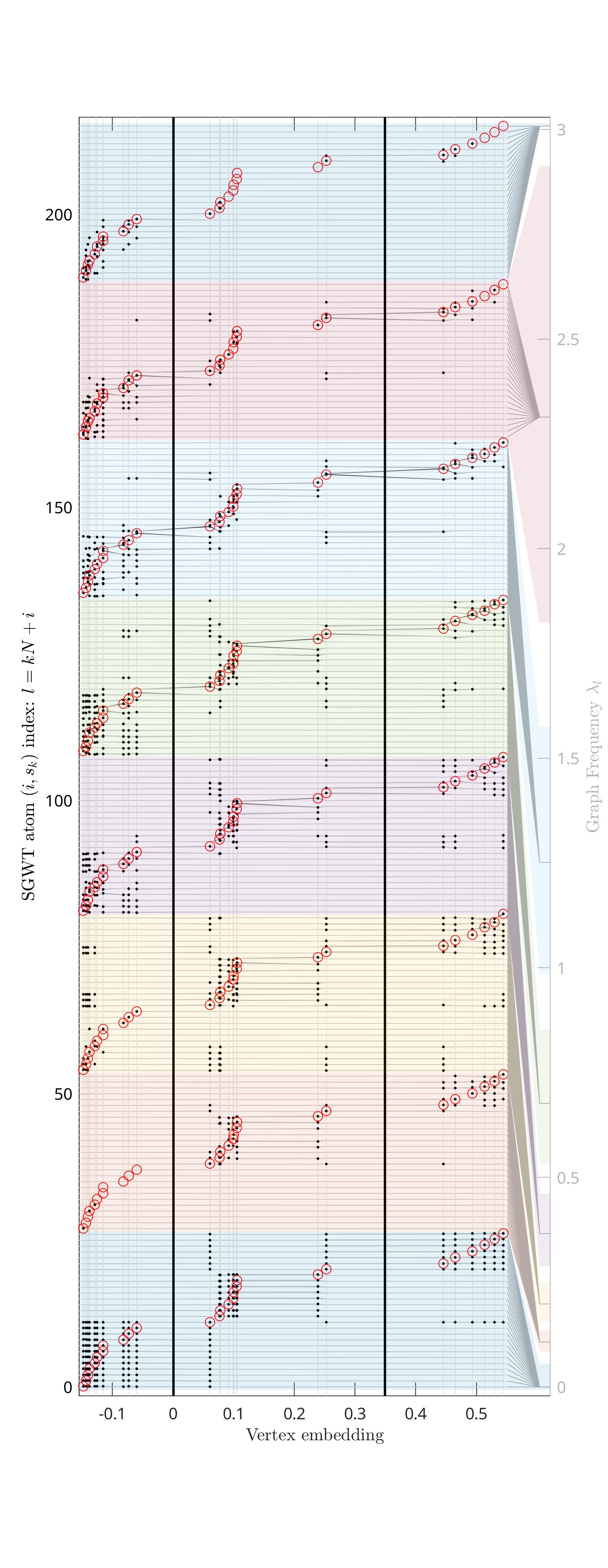}
        \caption{}\label{fig:sgwt_atoms:comb_lapl}
    \end{subfigure}
    \begin{subfigure}{0.49\linewidth}
        \centering
        \includegraphics[width=\sgwtwidth\linewidth,clip,trim=0 300 0 300]%
            {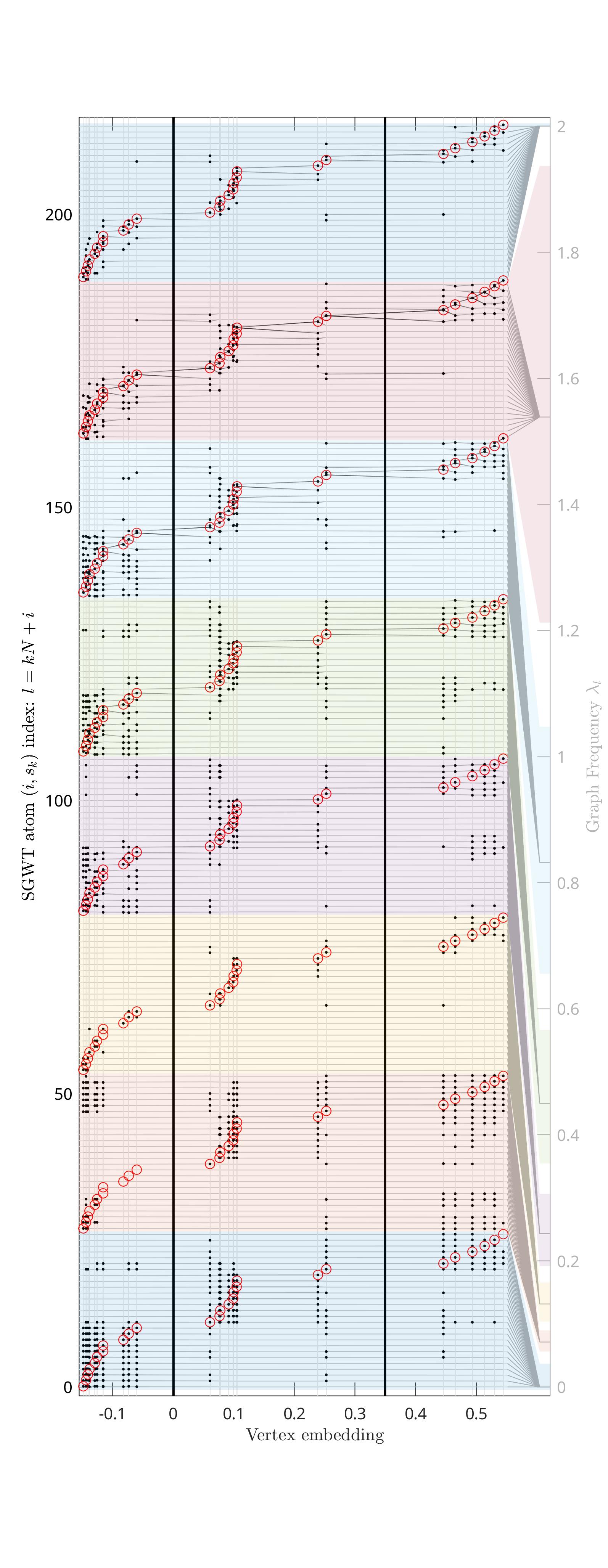}
        \caption{}\label{fig:sgwt_atoms:norm_lapl}
    \end{subfigure}
    
    \caption{SGWT atoms using 7 wavelet scales (and 1 scaling function) for \subref{fig:sgwt_atoms:comb_lapl} the  combinatorial Laplacian approach and \subref{fig:sgwt_atoms:norm_lapl} the normalized Laplacian approach. Atoms are sorted from bottom to top by scale first (coarsest to finest) and then by node embedding. Highlighted entries correspond to the $i^{\text{th}}$ node of the atom $g_{i,k}$. Each atom is mapped to the frequency of maximum spectral response of its associated bandpass filter $g_k(\lambda)$. The associated band is obtained by keeping all frequencies such that $g_k(\lambda)\geq 1.2$.}
    \label{fig:sgwt_atoms}
\end{figure*}

\def\sgwtwidth{1}
\begin{figure*}
    \centering
    \begin{subfigure}{0.49\linewidth}
        \centering
        \includegraphics[width=\sgwtwidth\linewidth,clip,trim=0 300 0 300]%
            {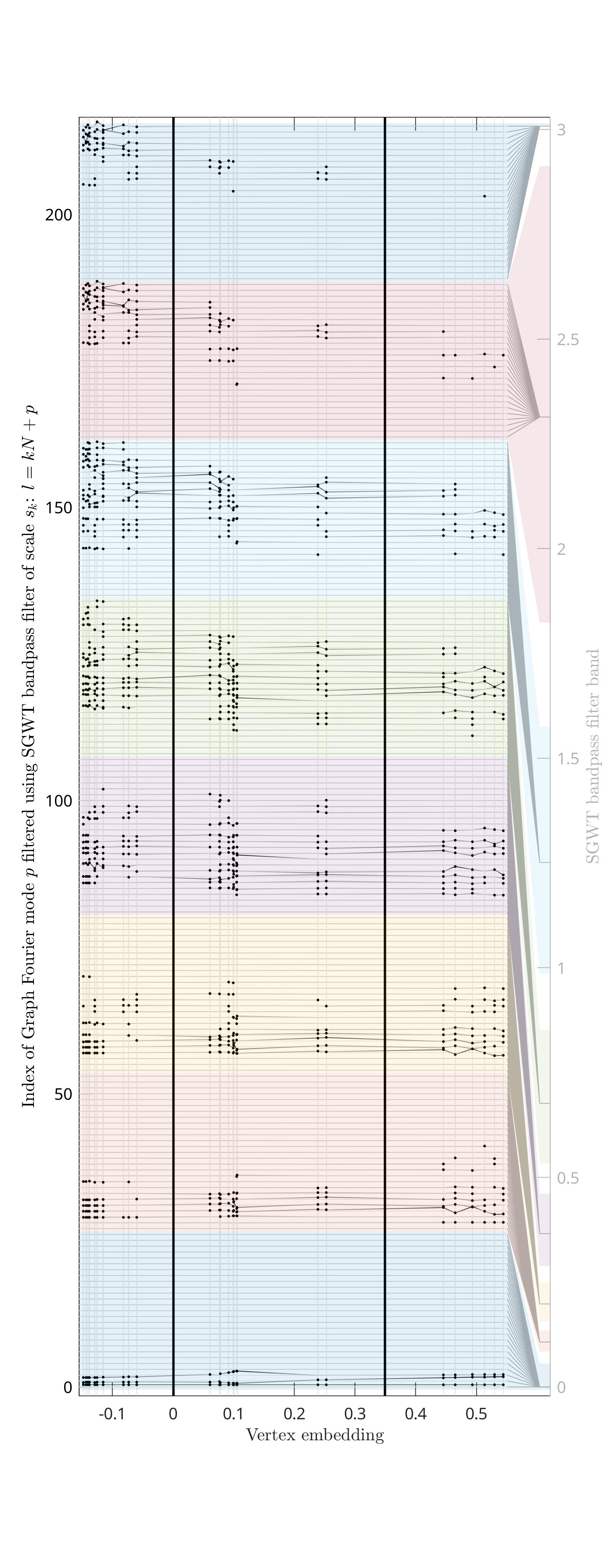}
        \caption{}\label{fig:sgwt_bands:comb_lapl}
    \end{subfigure}
    \begin{subfigure}{0.49\linewidth}
        \centering
        \includegraphics[width=\sgwtwidth\linewidth,clip,trim=0 300 0 300]%
            {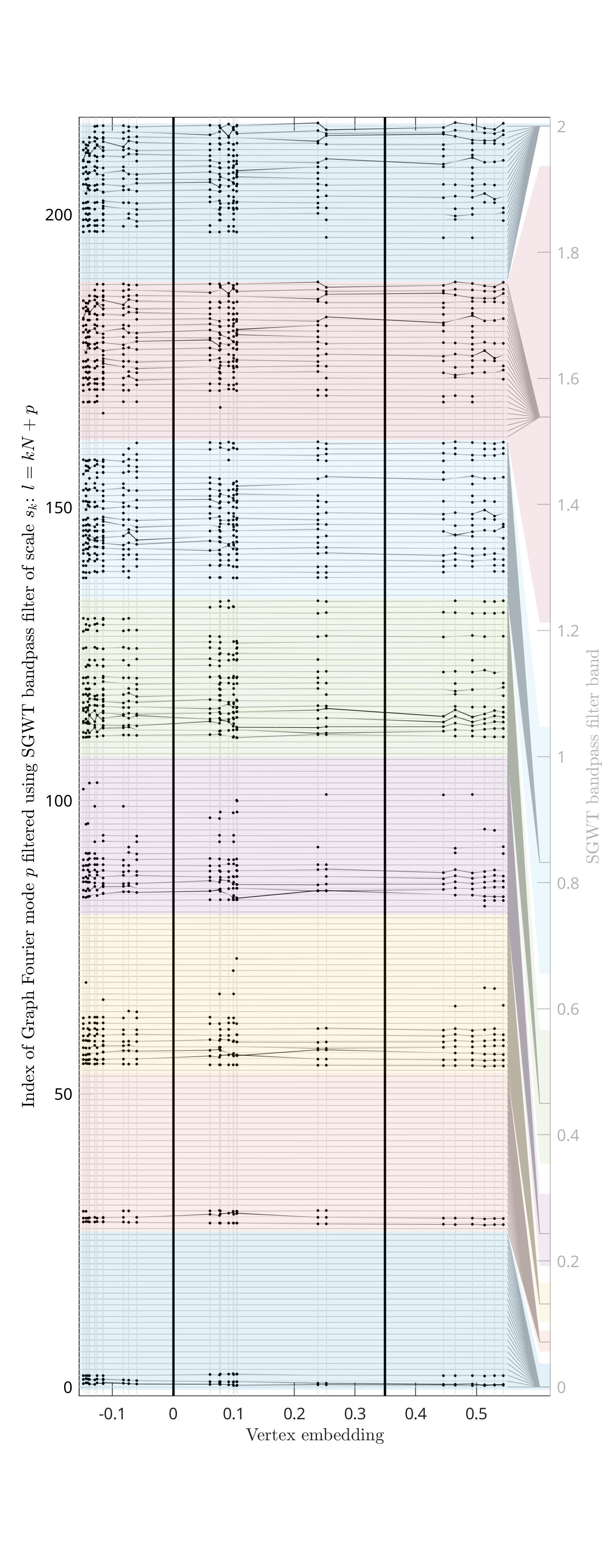}
        \caption{}\label{fig:sgwt_bands:norm_lapl}
    \end{subfigure}
    \caption{TODO.}
    \label{fig:sgwt_bands}
\end{figure*}

\section{Conclusion}
\label{sec:conclusion}
In this paper we have described a new set of visualization tools for graph signal processing, and their implementation in the \grasp Matlab toolbox. While it is well known that the GFT differs significantly from transforms for signals in regular domain, our goal is to encourage further research into this behavior, and into tools for graph signal representation that can be used in practice, by making it easier to visualize graph signal frequency information. 

\bibliographystyle{IEEEbib}
\bibliography{refs}

\vfill\pagebreak

\end{document}